\RequirePackage{lineno}
\documentclass[a4paper,11pt,epsfig]{article}

\usepackage{jheppub} 
                     
\usepackage[T1]{fontenc} 
\usepackage{graphicx,color,dcolumn,booktabs,bm}
\usepackage{longtable,lscape}
\usepackage{amsmath}
\usepackage{indentfirst}
\usepackage{epsfig}
\usepackage{epstopdf}   
\usepackage{slashed}  
\usepackage{color}
\usepackage{multirow}
\usepackage{soul}  
\usepackage{graphicx,color,dcolumn,booktabs,bm}
\usepackage{verbatim}
\usepackage{orcidlink}
\usepackage{appendix}

\newcommand{\BR}{{\mathcal B}}

\newcommand{\lum}{{\cal L}}

\newcommand{\xicp}{\Xi_{c}^{+}}
\newcommand{\xiz}{\Xi^{0}}

\newcommand{\piz}{\pi^{0}}

\newcommand{\ks}{K_{S}^{0}}
\newcommand{\sg}{\Sigma^{+}}
\newcommand{\kp}{K^{+}}
\newcommand{\xim}{\Xi^{-}}
\newcommand{\pim}{\pi^{-}}
\newcommand{\pip}{\pi^{+}}
\newcommand{\sgks}{\sg\ks}
\newcommand{\xipi}{\xiz\pip}
\newcommand{\xikp}{\xiz\kp}
\newcommand{\xipipi}{\xim\pip\pip}

\newcommand{\ep}{e^{+}}
\newcommand{\e}{e^{-}}
\newcommand{\ee}{\ep\e}
\newcommand{\eetocc}{\ee\to c \bar{c}}

\title{\boldmath Measurements of the branching fractions of $\Xi_{c}^{+}\to \Sigma^{+}K_{S}^{0}$, $\Xi_{c}^{+}\to \Xi^{0}\pi^{+}$, and $\Xi_{c}^{+}\to \Xi^{0}K^{+}$ at Belle and Belle II }

\preprint{\vbox{ \hbox{   }
\hbox{Belle II Preprint 2025-005 }
\hbox{KEK Preprint 2025-2}
}}

\collaboration{The Belle and Belle~II Collaborations}
\collaboration{The Belle and Belle II Collaborations}
  \author{I.~Adachi\,\orcidlink{0000-0003-2287-0173},} 
  \author{J.~K.~Ahn\,\orcidlink{0000-0002-5795-2243},} 
  \author{Y.~Ahn\,\orcidlink{0000-0001-6820-0576},} 
  \author{N.~Akopov\,\orcidlink{0000-0002-4425-2096},} 
  \author{S.~Alghamdi\,\orcidlink{0000-0001-7609-112X},} 
  \author{M.~Alhakami\,\orcidlink{0000-0002-2234-8628},} 
  \author{N.~Althubiti\,\orcidlink{0000-0003-1513-0409},} 
  \author{K.~Amos\,\orcidlink{0000-0003-1757-5620},} 
  \author{N.~Anh~Ky\,\orcidlink{0000-0003-0471-197X},} 
  \author{C.~Antonioli\,\orcidlink{0009-0003-9088-3811},} 
% ...

\abstract{Using 983.0~$\mathrm{fb}^{-1}$ and 427.9~$\mathrm{fb}^{-1}$ data samples collected with the Belle and Belle~II detectors at the KEKB and SuperKEKB asymmetric energy $e^+e^-$ colliders, respectively, we present studies of the Cabibbo-favored $\xicp$ decays ${\Xi_{c}^{+}\to \Sigma^{+}K_{S}^{0}}$ and $\Xi_{c}^{+}\to \Xi^{0}\pi^{+}$, and the singly Cabibbo-suppressed decay $\Xi_{c}^{+}\to \Xi^{0}K^{+}$. The ratios of branching fractions of ${\Xi_{c}^{+}\to \Sigma^{+}K_{S}^{0}}$ and $\Xi_{c}^{+}\to \Xi^{0}K^{+}$ relative to that of $\Xi_{c}^{+}\to\Xi^{-}\pi^{+}\pi^{+}$ are measured for the first time, while the ratio ${\mathcal B}(\Xi_{c}^{+}\to\Xi^{0}\pi^{+})/\BR(\Xi_{c}^{+}\to\Xi^{-}\pi^{+}\pi^{+}) $ is also determined and improved by an order of magnitude in precision. The measured branching fraction ratios are
\begin{align*}
\frac{\mathcal{B}(\Xi_{c}^{+} \to \Sigma^{+}K_{S}^{0})}{\mathcal{B}(\Xi_{c}^{+}\to \Xi^{-}\pi^{+}\pip)}&= 0.067 \pm 0.007 \pm 0.003  , \\
\frac{\mathcal{B}(\Xi_c^{+} \to \Xi^{0}\pi^{+})}{\mathcal{B}(\Xi_{c}^{+}\to \Xi^{-}\pi^{+}\pip)} &=  0.251 \pm 0.005  \pm 0.010, \\
\frac{\mathcal{B}(\Xi_c^{+} \to \Xi^{0}K^{+})}{\mathcal{B}(\Xi_{c}^{+}\to \Xi^{-}\pi^{+}\pip)} &= 0.017 \pm 0.003  \pm 0.001 .
\end{align*}
Additionally, the ratio ${\mathcal B}(\Xi_{c}^{+}\to\Xi^{0}K^{+})/{\mathcal B}(\Xi_{c}^{+}\to\Xi^{0}\pi^{+})$ is measured to be $ 0.068 \pm 0.010  \pm 0.004$. Here, the first and second uncertainties are statistical and systematic, respectively.
Multiplying the ratios by the branching fraction of the normalization mode, ${\mathcal B}(\Xi_{c}^{+}\to\Xi^{-}\pi^{+}\pip)= (2.9\pm 1.3)\%$, we obtain the following absolute branching fractions 
\begin{align*}
{\mathcal B}(\Xi_{c}^{+}\to\Sigma^{+}K^{0}_{S}) &= (0.194 \pm 0.021  \pm 0.009  \pm 0.087 )\%, \\
{\mathcal B}(\Xi_{c}^{+}\to\Xi^{0}\pi^{+}) &=  (0.728 \pm 0.014  \pm 0.027  \pm 0.326 )\%, \\
{\mathcal B}(\Xi_{c}^{+}\to\Xi^{0}K^{+}) &= (0.049 \pm 0.007   \pm 0.003 \pm 0.022 )\%,
\end{align*}
where the third uncertainties are from $\BR(\Xi_{c}^{+}\to\Xi^{-}\pi^{+}\pi^{+})$.

}

\begin{document}
\maketitle
\flushbottom

\section{Introduction}
\noindent Charmed baryon decays exhibit rich experimental phenomena, and provide a valuable probe of the non-perturbative dynamics of quantum chromodynamics. The decay amplitudes generally receive contributions from both factorizable and non-factorizable effects.
The latter arising from internal $W$-emission and $W$-exchange quark-level processes play a pivotal role, introducing significant challenges for the theoretical predictions in hadronic weak decays of charmed baryons~\cite{charmedBaryon2022}. 
Extensive efforts have been devoted to studying charmed baryon decays, aiming to develop a theoretical model for calculating non-factorizable contributions.

Measurements of the branching fractions of charmed baryons are indispensable for theoretical analyses.
Although Belle reported the absolute branching fraction of $\Xi_{c}^{+}\to\Xi^{-}\pi^{+}\pi^{+}$ as $(2.9\pm1.3)$\% in 2019~\cite{xicpabsbf2019}, only a few other $\Xi^{+}_{c}$ decays have been measured~\cite{xicpabsbf2019,xicp1bf2024}. Theoretical calculations for the two-body hadronic weak decays of the $\Xi^{+}_{c}$ have been performed using a dynamical model~\cite{theory10poleca2020} and $\rm SU(3)$ flavor [${\rm SU(3)}_{f}$] symmetry methods~\cite{theory8su3f2019,theory9su3f2020,theory12su3f2022,theory11su3f2022,theory14su3f2023,theory100su3f2024,theory13su3f2023}. 
Nevertheless, the predictions of these two approaches for Cabibbo-favored~(CF) and singly Cabibbo-suppressed (SCS) decay modes are not fully consistent with each other.
One such mode is the CF decay $\xicp \to \Sigma^{+}K_{S}^{0}$, with a predicted branching fraction around $10^{-2}$~\cite{theory10poleca2020,theory8su3f2019,theory9su3f2020,theory12su3f2022,theory11su3f2022,theory14su3f2023,theory100su3f2024,theory13su3f2023}, which was not yet measured. Another CF decay, $\Xi_{c}^{+} \to \Xi^{0} \pi^{+}$, has been measured, yet its branching fraction had a large uncertainty~\cite{exp1}. Meanwhile, the branching fraction of the SCS decay $\Xi_{c}^{+}\to \Xi^{0}K^{+}$ is predicted to be in the range of $10^{-3}$ to $10^{-2}$ by several theoretical models~\cite{theory10poleca2020,theory8su3f2019,theory9su3f2020,theory12su3f2022,theory11su3f2022,theory14su3f2023,theory100su3f2024,theory13su3f2023}, but has never been observed. Exploiting synergies with the analysis of $\Xi_{c}^{+}\to \Xi^{0}\pip$, we search for the SCS $\xicp$ decay for the first time in Belle and Belle II data samples.

We present the first measurements of the branching fractions for ${\Xi_c^+ \to \sg \ks}$ and $\Xi_c^+ \to \Xi^0 K^+$ decays and an improved measurement of the branching fraction of ${\Xi_c^+ \to \Xi^0 \pi^+}$. 
The $\Xi_c^+ \to \Xi^{-} \pip \pip$ decay is taken as the normalization mode for these measurements. This analysis uses data samples with integrated luminosities of 983.0~$\mathrm{fb}^{-1}$~\cite{lumB1} and 427.9~$\mathrm{fb}^{-1}$~\cite{lumB2} collected with the Belle and Belle~II detectors operating at the KEKB and SuperKEKB asymmetric-energy $\ee$ colliders, respectively.
Charge-conjugate modes are implied throughout the paper.

\section{Belle and Belle~II experiments}

The Belle detector~\cite{lumB1,Belle1} operated from 1999 to 2010 at the $e^+e^-$ interaction point~(IP) of the KEKB asymmetric-energy $\ee$ collider~\cite{KEKB1,KEKB2}.
Belle was a large solid-angle, cylindrical magnetic spectrometer that consisted of a silicon vertex detector, a central drift chamber, an array of aerogel threshold Cherenkov counters, a barrel-like arrangement of time-of-flight scintillation counters, an electromagnetic calorimeter (ECL) comprised of CsI(Tl) crystals located inside a superconducting solenoid coil that provided a $1.5~\hbox{T}$ axial magnetic field, and an iron flux return placed outside the coil, instrumented with resistive-plate chambers to detect $K^{0}_{L}$ mesons and to identify muons.
A detailed description of the detector can be found in refs.~\cite{lumB1,Belle1}. 

The Belle~II detector~\cite{BelleII} is located at the IP of the SuperKEKB asymmetric-energy $\ee$ collider~\cite{superKEKB}, the upgraded successor of KEKB.
Belle~II is an upgraded version of the Belle detector and consists of several new subsystems and substantial upgrades to others. The new vertex detector includes two inner layers of
pixel sensors and four outer layers of double-sided silicon micro-strip sensors.
For the data sample used in this analysis, the second-pixel layer was incomplete, covering only one-sixth of the azimuthal angle.
A new central drift chamber surrounding the vertex detector is used to measure the momenta and electric charges of charged particles. 
A time-of-propagation detector in the barrel and an aerogel ring-imaging Cherenkov detector in the forward endcap provide information for the identification of charged particles, supplemented by ionization energy loss measurements in the central drift chamber.
The ECL readout electronics have been upgraded to cope with the higher beam-induced background environment at Belle II. 
The superconducting solenoid coil and the iron flux return for Belle are reused in Belle II, with the inner two layers of the barrel and the endcap resistive plate chambers of the $K^{0}_{L}-$muon detector replaced by plastic scintillator modules.
The $+z$ axis of the laboratory frame is defined as the central solenoid axis in the direction of the $e^-$ beam, common to Belle and Belle II.


\section{Data samples \label{datasets}}
The data recorded at center-of-mass (c.m.) energies at or near the $\Upsilon(1S)$, $\Upsilon(2S)$, $\Upsilon(3S)$, $\Upsilon(4S)$, and $\Upsilon(5S)$ resonances by the Belle detector, and at or near the $\Upsilon(4S)$ and 10.75~GeV by the Belle II detector are used in this analysis.
The data samples have integrated luminosities of 983.0~$\mathrm{fb}^{-1}$~\cite{lumB1} and 427.9~$\mathrm{fb}^{-1}$~\cite{lumB2} for Belle and Belle~II, respectively.

Monte Carlo (MC) samples of simulated events are generated using {\sc evtgen}~\cite{evtgen} and used to optimize the $\xicp$ candidate selection criteria and determine the reconstruction efficiency.
We generate continuum $\eetocc$ with {\sc pythia6}~\cite{pythia1} for Belle and {\sc kkmc}~\cite{kkmc} and {\sc pythia8}~\cite{pythia2} for Belle II, where one of the charm quarks is required to hadronize into a $\xicp$ baryon.
Simulated $\xicp$ decays into $\sgks$, $\xipi$, $\xikp$, and $\xipipi$ are then generated with a phase space model. 
To optimize the selection criteria and study possible backgrounds, the inclusive MC samples of $\Upsilon(4S) \to B\bar{B}$ decays at Belle and Belle II, as well as $\Upsilon(1S,2S,3S)$ decays and $\Upsilon(5S) \to B_{(s)}^{(*)}\bar{B}_{(s)}^{(*)}$ decays at Belle, are generated using the same packages.~The continuum background from $e^+e^- \to q\bar{q}$ processes, where $q$ indicates a $u$, $d$, $c$, or $s$ quark, is generated with {\sc pythia} for hadronization and {\sc evtgen} for subsequent decays of hadrons.
Final state radiation effects are accounted for using the {\sc PHOTOS} package~\cite{photos}.
Simulation of the detector response uses the {\sc geant3}~\cite{geant3} and {\sc geant4}~\cite{geant4} software packages for Belle and Belle II, respectively.

\section{Selection criteria}

We reconstruct the decays $\xicp\to\sgks$, $\xipi$, $\xikp$, and $\xipipi$, with subsequent decays $\xiz\to\Lambda\piz$, $\xim\to\Lambda\pim$, $\Lambda\to p\pim$, $\sg\to p\piz$, and $\ks\to\pip\pim$.
The Belle~II software~\cite{basf2} is used for analyzing the data of both experiments, taking advantage of software improvements in Belle~II. 
The Belle data is converted to the Belle II data format using the B2BII software package~\cite{b2bii}. To improve the sensitivity for signal candidates, the selection criteria described below are optimized by maximizing the figure-of-merit, defined as $N_{\rm sig}/\sqrt{N_{\rm sig}+N_{\rm bkg}}$. 
Here, $N_{\rm sig}$ is the expected signal yield based on either theoretical predictions or previous measurements~\cite{theory10poleca2020,pdg}, and $N_{\rm bkg}$ is the background yield determined from the inclusive MC samples in the $\Xi_c^+$ signal regions and scaled by the ratio of yields between data and inclusive MC in the normalized $\xicp$ sideband regions. The $\Xi_c^+$ signal regions are defined as $|M(\Sigma^{+}K_{S}^{0})-m_{\xicp}| < $ 30 MeV/$c^{2}$ and $|M(\Xi^{0} \pip/\Xi^{0}K^+)-m_{\xicp}| < $ 20 MeV/$c^{2}$. The ranges of the $\Xi_c^+$ signal regions are approximately three times the experimental invariant mass resolution ($\sigma$). The $\xicp$ sideband regions are defined as 50 MeV/$c^{2}$~$<|M(\Sigma^{+}K_{S}^{0})-m_{\xicp}| < $ 80 MeV/$c^{2}$ and 35 MeV/$c^{2}$~$<|M(\Xi^{0} \pip/\Xi^{0}K^+)-m_{\xicp}| < $ 55 MeV/$c^{2}$.
Here and throughout this paper, $m_{i}$ denotes the known mass of the particle $i$~\cite{pdg} and $M(jk)$ indicates the invariant mass of the reconstructed particles $j$ and $k$.


The impact parameters of charged tracks, except for those originating from $\ks$ or hyperon ($\xim$, $\sg$, or $\Lambda$) decays, which are the distances of the closest approach of the reconstructed trajectory perpendicular to and along the $z$-axis to the centroid of the calibrated IP region, are required to be less than 1.0~cm and 3.0~cm, respectively. These impact parameter requirements help to suppress misreconstructed tracks and beam backgrounds.
The identification of charged tracks is achieved with the likelihood ratio $\mathcal{R}(h|h') = \mathcal{L}(h)/ [\mathcal{L}(h)+ \mathcal{L}(h')]$, where $\mathcal{L}(h^{(\prime)})$ represents the likelihood of the charged track being a hadron ($h^{(\prime)} = \pi$, $K$, or $p$). The likelihood is determined using a Particle Identification (PID) algorithm, which integrates information from the Belle or Belle II subdetectors~\cite{BellePID1}.
For pions, we require $\mathcal{R}(\pi|K)>0.6$; for kaons, we require $\mathcal{R}(K|\pi)>0.6$. 
These PID requirements have reconstruction efficiencies in the range of 88\%–92\%.
To suppress backgrounds from low-momentum pions and kaons, we require the momentum in the laboratory frame of pions and kaons to be greater than 0.5~GeV/$c$ for $\xicp\to\xiz\pip$ and $\xicp\to\xiz\kp$, respectively.

In the reconstruction of $\Xi^-\to \Lambda\pi^-$ and $\Xi^0\to \Lambda\pi^0$, the selected $\Lambda$ candidates are combined with a $\pim$ or $\piz$. The selections of $\xim$ and $\xiz$ candidates follow those in Ref.~\cite{omega2018}. A vertex fit is performed to the entire $\Xi^{-(0)}$ decay chain, with the $p\pi^-$ and diphoton masses constrained to match the known $\Lambda$ and $\pi^0$ masses~\cite{pdg}. The distance of the $\xim$ decay vertex to the IP is required to be greater than 0.1 cm, whereas the displacement of the $\xiz$ vertex must be greater than 1.4 cm. The selected $\Xi^{-(0)}$ candidate satisfies the requirement $\cos\alpha>$ 0, where $\alpha$ is the angle between the momentum and the vertex vector of $\Xi^{-(0)}$. The vertex vector points from the IP to the fitted $\Xi^{-(0)}$ vertex.
The $\Lambda$ candidates are reconstructed via the $\Lambda \to p \pi^-$ decay. The distance between the decay vertex of each $\Lambda$ candidate and the IP is required to be greater than 0.35 cm~\cite{xicp1bf2024}.
The signal region of $\Lambda$ candidates is defined as $|M(p\pi^{-})-m_{\Lambda}| < $ 3.5 MeV/$c^{2}$ (about $3 \sigma$).
Each $\pi^-$ candidate from the $\Xi^-$ decay is required to have a transverse momentum greater than 0.1~GeV/$c$ to remove backgrounds from low-momentum pions.
An ECL cluster is used to form a photon candidate if it is not associated with the extrapolation of a charged track. Candidate $\pi^0$'s from $\xiz$ decays are reconstructed from pairs of photons selected from energy deposits in the ECL.
To suppress low-momentum and fake photons, each photon candidate is required to have an energy greater than 30~MeV in the ECL barrel region ($-0.63<\cos\theta<0.85$); 
50 (80)~MeV for Belle (Belle~II) in the forward endcap ($0.85<\cos\theta<0.98$);
and 50 (60)~MeV in the backward endcap ($-0.91<\cos\theta<-0.63$), where $\theta$ is the polar angle in the laboratory frame. The signal region of the $\piz$ candidates is defined as $|M({\gamma\gamma})-m_{\piz}| < $ 11.6~MeV/$c^{2}$ (about $2 \sigma$).
In addition, the momenta of the $\pi^0$ candidates in the laboratory frame are required to exceed 0.25~GeV/$c$. 
Finally, $\xim$ and $\xiz$ candidates are formed from $\Lambda\pim$ and $\Lambda\piz$ combinations, respectively.
The signal regions of $\xim$ and $\xiz$ candidates are defined as $|M({\Lambda\pi^{-}})-m_{\xim}| < $ 6.6 MeV/$c^{2}$ (about $ 3 \sigma$) and $|M(\Lambda\pi^{0})-m_{\xiz}| < $ 6.0 MeV/$c^{2}$ (about $2 \sigma$), respectively.

The $\ks$ candidates decaying to $\pip\pim$ are selected with an artificial neural network for Belle~\cite{Belleks1,Belleks2}.
For Belle~II, the flight distance of each $K_{S}^{0}$ is required to be larger than 0.5~cm to remove random combinations of pions. 
The flight distance is calculated as the projection of the vector that joins the IP to the decay vertex onto the direction of its momentum.
The signal region is defined as $|M(\pip\pim) - m_{\ks}| < 7.0$ MeV/$c^{2}$ (about $2 \sigma$).
The $\Sigma^{+}$ candidates are reconstructed in the $\Sigma^{+} \to p\pi^{0}$ decay mode, with the likelihood ratio requirements $\mathcal{R}(p|\pi)>0.6$ and $\mathcal{R}(p|K)>0.6$ for the proton candidates.
The distance of the closest approach of the proton to the IP in the transverse plane is greater than $0.1$~cm.
The photon energy requirements on the $\piz$ from the $\sg$ are the same as those for the $\piz$ from the $\xiz$. A loose signal region is used to select $\piz$ candidates with a range of (0.110, 0.155) GeV/$c^{2}$.
To reduce the combinatorial backgrounds, the $\pi^0$ momentum in the $e^+e^-$ c.m.\ frame must be greater than 0.45 GeV/$c$. 
A vertex fit is applied to the entire $\sg$ decay chain, with a $\piz$ invariant mass constraint.
The requirement $\cos\alpha>$ 0 is also applied to select $\sg$ candidates. The efficiency loss for this selection is less than 1\%. For selected $\sg$ candidates, $|M(p\pi^{0})-m_{\Sigma^{+}}|< 7.0$ MeV/$c^2$ (about $1.5 \sigma$) is required.

For the $\Xi_{c}^{+}$ candidates,
a vertex fit is applied for each mode using the $\texttt{TreeFit}$ algorithm with mass constraints for the intermediate states and the constraint that the $\xicp$ originates from the interaction region~\cite{b2fit}. 
To suppress backgrounds, especially those from $B$-meson decays, we require the scaled momentum ${x}_{p} =p_{\Xi_{c}^{+}}^{*}/p_{\rm max}$ to be larger than 0.55. Here, the $p_{\Xi_{c}^{+}}^{*}$ is the momentum of the $\xicp$ candidate in the $e^+e^-$ c.m.\ frame, and $p_{\rm max}$ = $\frac{1}{c}\sqrt{E_{\rm beam}^{2}-M^2_{\Xi_c^{+}}{c}^4}$, where the $E_{\rm beam}$ is the beam energy in the $e^+e^-$ c.m.\ frame and $M_{\Xi_c^{+}}$ is the mass of the $\xicp$ candidate.

\section{Measurements of the branching fractions}

\begin{figure}[htbp]
	\centering
	\includegraphics[width=7cm]{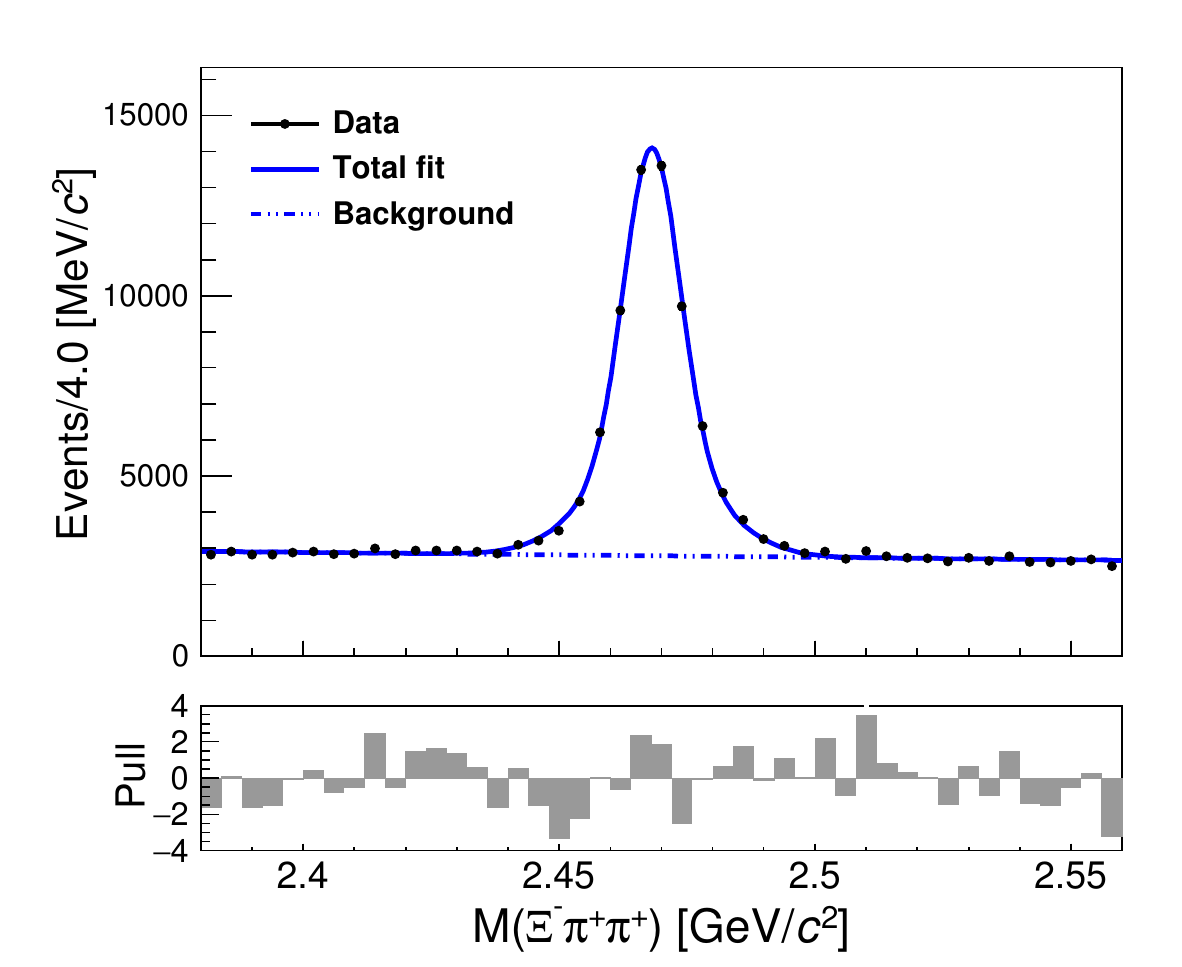}\put(-60,130){\bf (a)}
	\includegraphics[width=7cm]{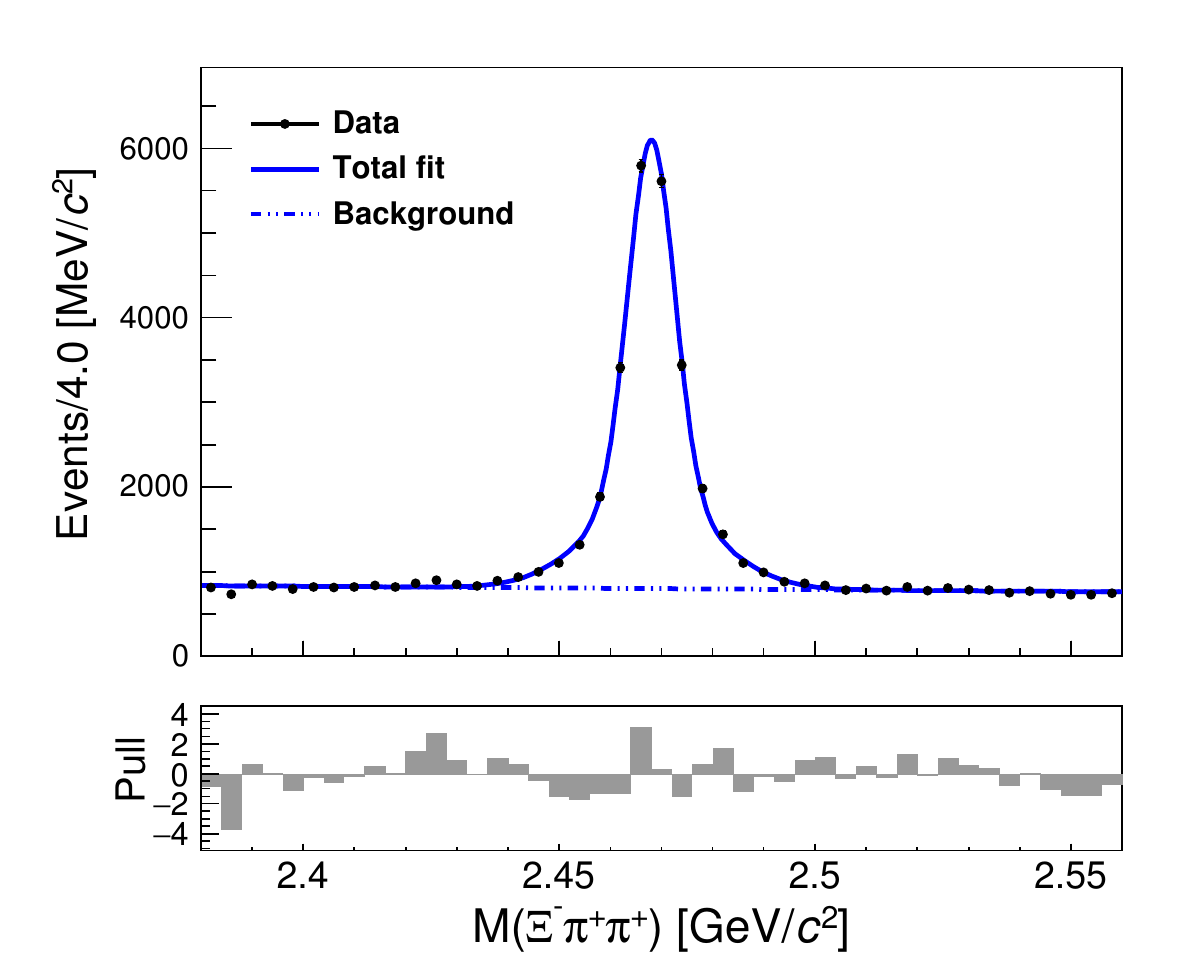}\put(-60,130){\bf (b)}
	\put(-355,155){\scriptsize Belle} \put(-330,155){\scriptsize $\int \lum dt$ = 983.0~fb$^{-1}$}
	\put(-155,155){\scriptsize Belle~II} \put(-120,155){\scriptsize $\int \lum dt$ = 427.9~fb$^{-1}$}
	\caption{Invariant mass distributions of $\Xi^{-} \pip \pip$ from (a) Belle and (b) Belle II data. The markers with error bars represent the data, the solid blue curves show the fit results, and the dashed blue curves show the background component of the fit. }
	\label{mximpippip-data}
\end{figure}

The $\xim\pip\pip$ invariant mass distributions for $\xicp$ candidates in Belle and Belle II data after applying all event selection criteria are shown in the figure~\ref{mximpippip-data}, together with the results of an unbinned extended maximum-likelihood (EML) fit.
A sum of two Gaussians is used as the signal probability density function (PDF) for the $\Xi_c^{+}$ shape, while the background is represented by a first-order polynomial.
All signal and background parameters are allowed to float in the fit. The distributions of pull $= (N_{\rm data}-N_{\rm fit})$/$\sqrt{N_{\rm data}}$ are also displayed in figure~\ref{mximpippip-data}, where $N_{\rm data}$ is the number of entries in each bin from data, and $N_{\rm fit}$ is the fit result in each bin.
The fitted signal yields are summarized in table~\ref{Nxicp}.

\begin{figure}[htbp]
	\centering
	\includegraphics[width=7cm]{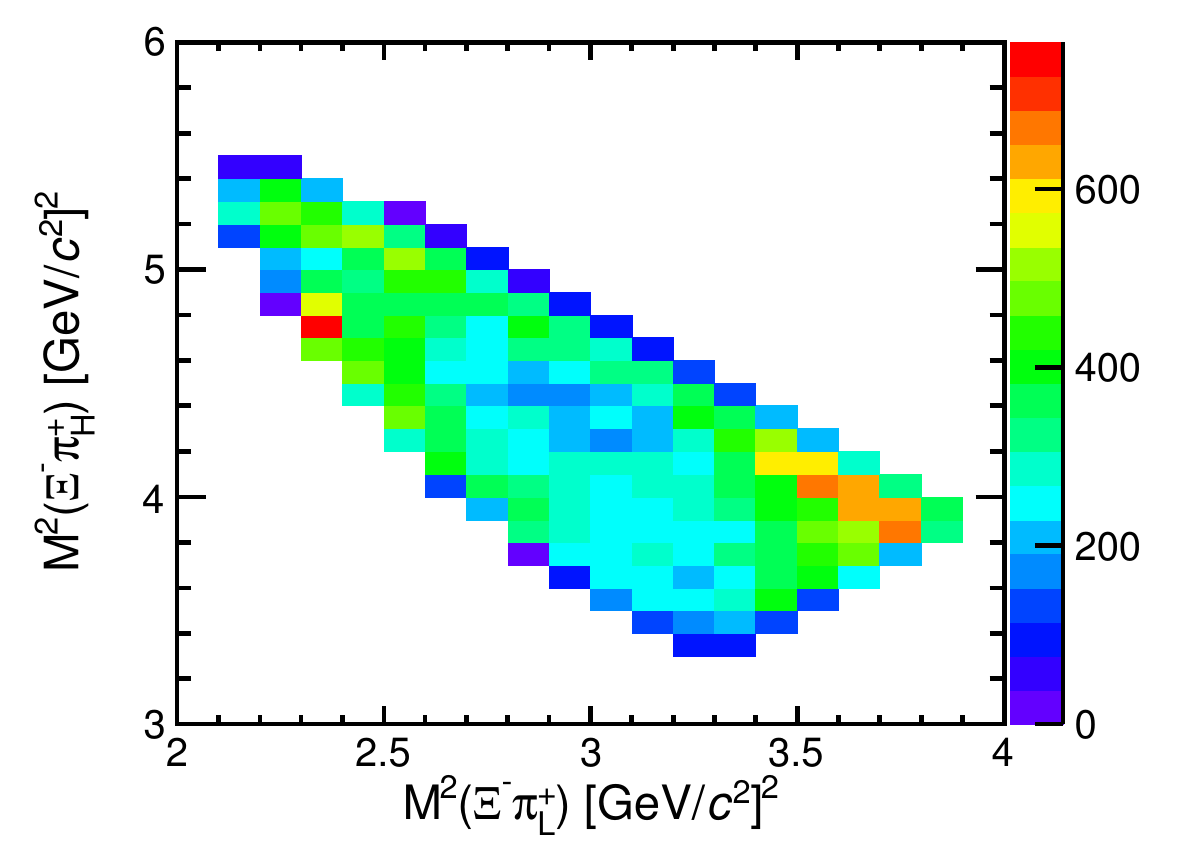}\put(-60,110){\bf (a)}
	\includegraphics[width=7cm]{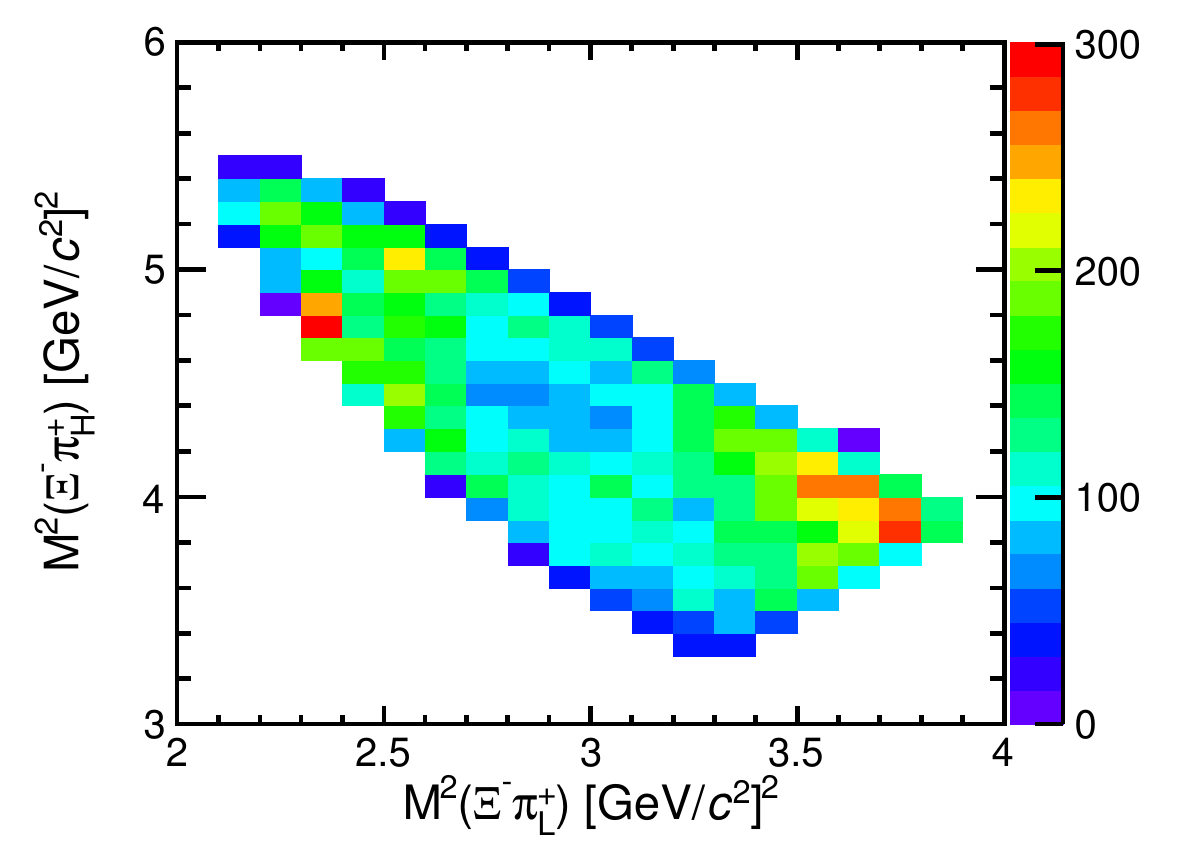}\put(-60,110){\bf (b)}
	\put(-360,140){\scriptsize Belle} \put(-335,140){\scriptsize $\int \lum dt$ = 983.0~fb$^{-1}$}
	\put(-160,140){\scriptsize Belle~II} \put(-125,140){\scriptsize $\int \lum dt$ = 427.9~fb$^{-1}$}
	\caption{Dalitz distributions of the reconstructed $\Xi^{-} \pip \pip$ candidates from (a) Belle and (b) Belle II data in the $\xicp$ signal region with the normalized $\xicp$ sideband events subtracted. }
	\label{mximpippip-dalitz}
\end{figure}

The reconstruction efficiency of the three-body decay $\xicp\to\xipipi$ can vary across the phase space.
Figure~\ref{mximpippip-dalitz} shows the Dalitz distributions~\cite{dalitz} of $M^{2}(\Xi^{-}\pi_{L}^{+})$ versus $M^{2}(\Xi^{-}\pi_{H}^{+})$ in the $\xicp$ signal region after subtracting the normalized events from $\xicp$ sideband regions, where the $\Xi^-\pi^+$ combination with a higher (lower) invariant mass is labeled as $M^{2}(\Xi^{-}\pi_{H}^{+})$ ($M^{2}(\Xi^{-}\pi_{L}^{+})$).
Here, the $\xicp$ signal and sideband regions for $\xicp\to \xipipi$ mode are defined as $|M(\Xi^{-}\pip\pip)-m_{\xicp}| < $ 24 MeV/$c^{2}$ (about $3 \sigma$) and 40 MeV/$c^{2}$$<|M(\Xi^{-}\pip\pip)-m_{\xicp}|<64$ MeV/$c^{2}$, respectively.
We divide the Dalitz plot into 20 $\times$ 30 bins and then apply a bin-by-bin correction for the efficiency.
The reconstruction efficiency averaged over the Dalitz plot is determined via the formula 
\begin{equation}
\label{xpop}
		\varepsilon
	    = \frac{ \sum_{i}N_{i} } {\sum_{i}(N_{i}/\varepsilon_{i})}, \\
\end{equation}
where $i$ is summed over all bins; the reconstruction efficiency of each Dalitz plot bin ($\varepsilon_{i}$) is obtained from MC simulation. 
Here, $N_{i}$ is the number of signal candidates for the $i^{\rm th}$ bin in data calculated as $N_{i} = N_{i}^{\rm tot} - N_{\rm SR}^{\rm bkg}f_{i}^{\rm bkg}$, where $N_{i}^{\rm tot}$ is the yield in the $i^{\rm th}$ bin of the Dalitz distribution in the $\xicp$ signal region, $N^{\rm bkg}_{\rm SR}$ is the fitted background yield in the $\xicp$ signal region in data, and $f_{i}^{\rm bkg}$ is the fraction of background in the $i^{\rm th}$ bin, with $\sum_{i}f_{i}^{\rm bkg} = 1$. These fractions are obtained from the Dalitz plot of events in the normalized $\xicp$ sideband regions in data~\cite{xicp1bf2024}.
The signal efficiencies $\varepsilon$ are determined to be ($11.37\pm 0.02$)\% and ($10.98\pm 0.02$)\% for Belle and Belle~II, respectively, and are listed in table~\ref{Nxicp}. 

\begin{figure}[htbp]	
	\centering
	\includegraphics[width=7.0cm]{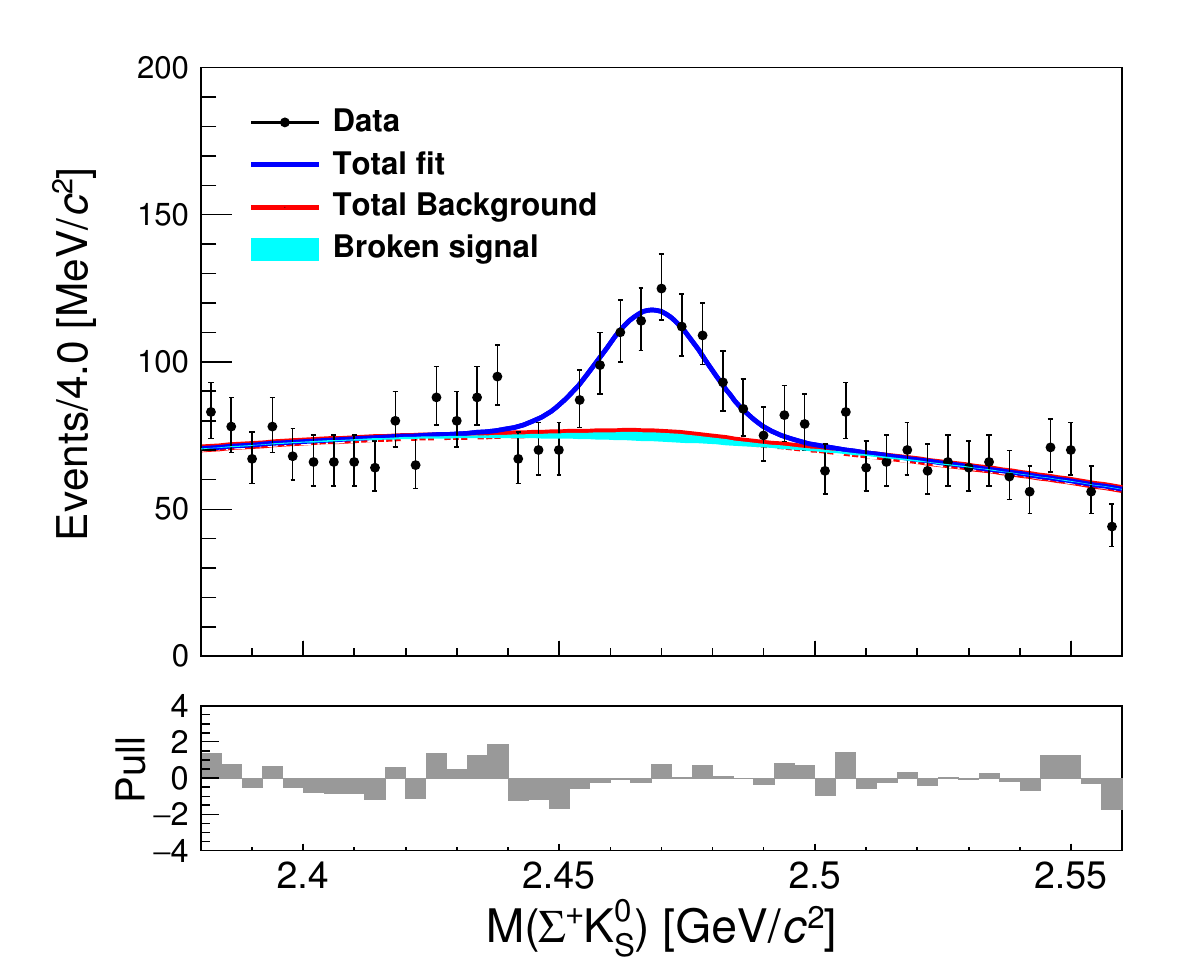}	
    \put(-60,130){\bf (a)}		
	\includegraphics[width=7.0cm]{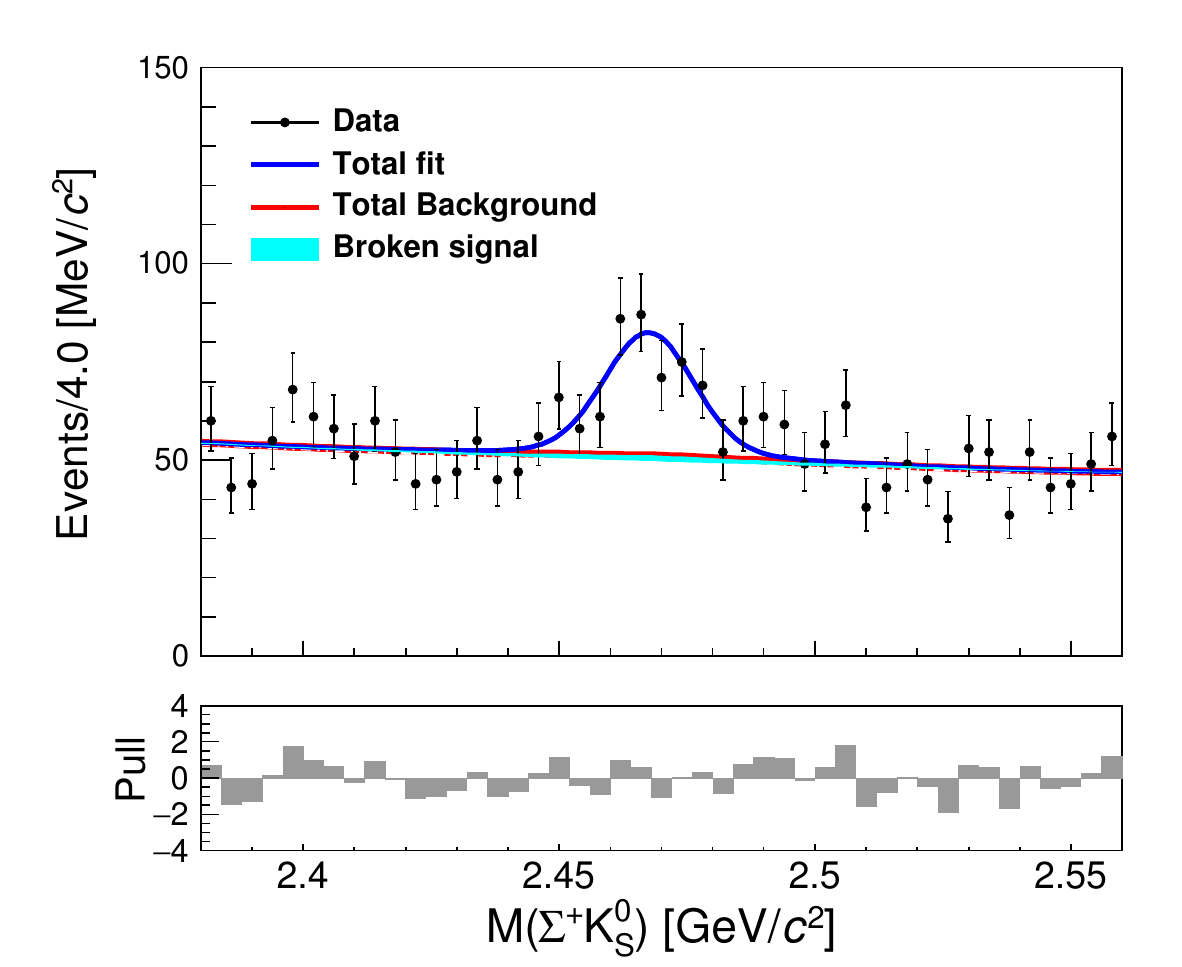}	
    \put(-60,130){\bf (b)}		
	\put(-355,155){\scriptsize Belle} \put(-330,155){\scriptsize $\int \lum dt$ = 983.0~fb$^{-1}$}
	\put(-155,155){\scriptsize Belle~II} \put(-120,155){\scriptsize $\int \lum dt$ = 427.9~fb$^{-1}$}
\vspace{0.5cm}
	\includegraphics[width=7.0cm]{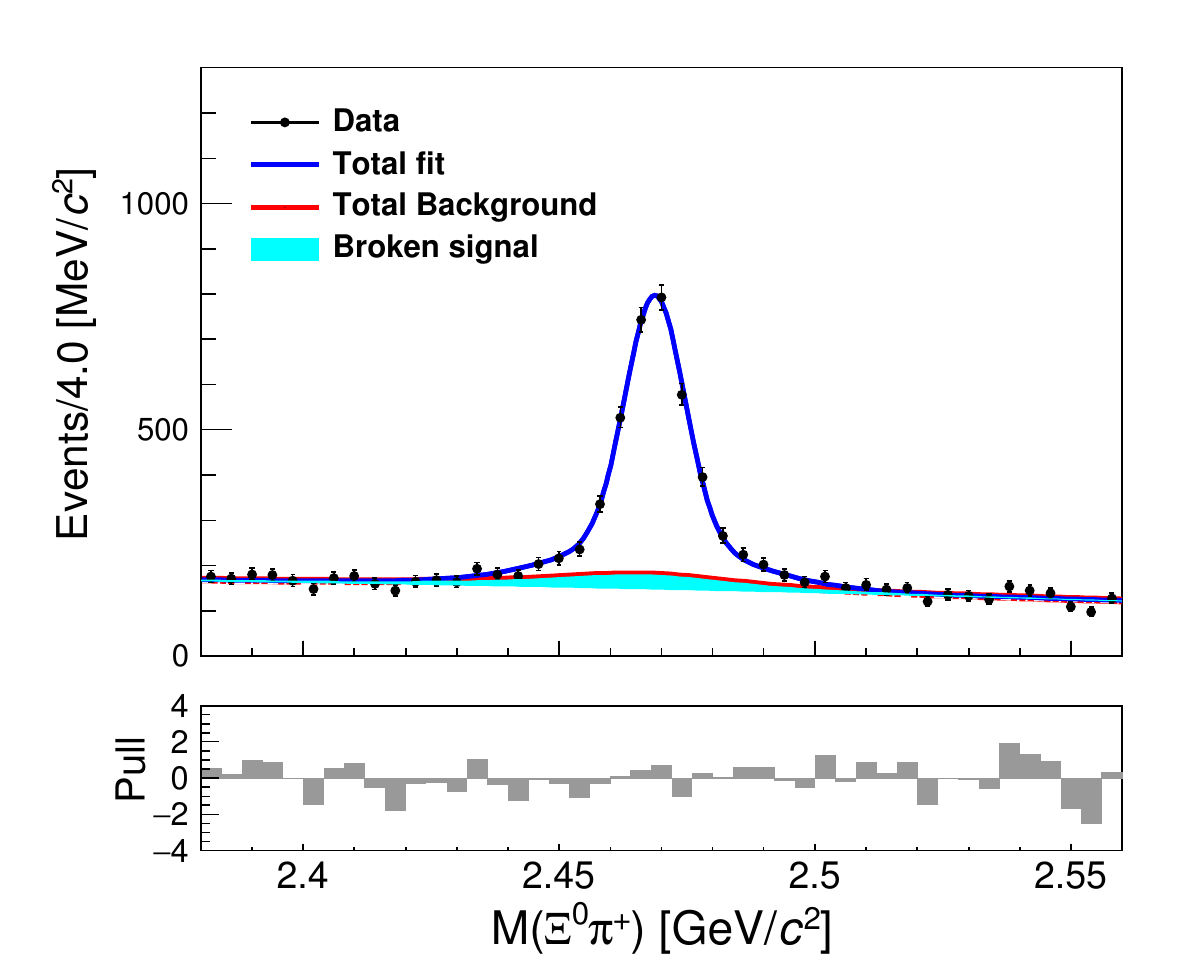}	\put(-60,130){\bf (c)}
	\includegraphics[width=7.0cm]{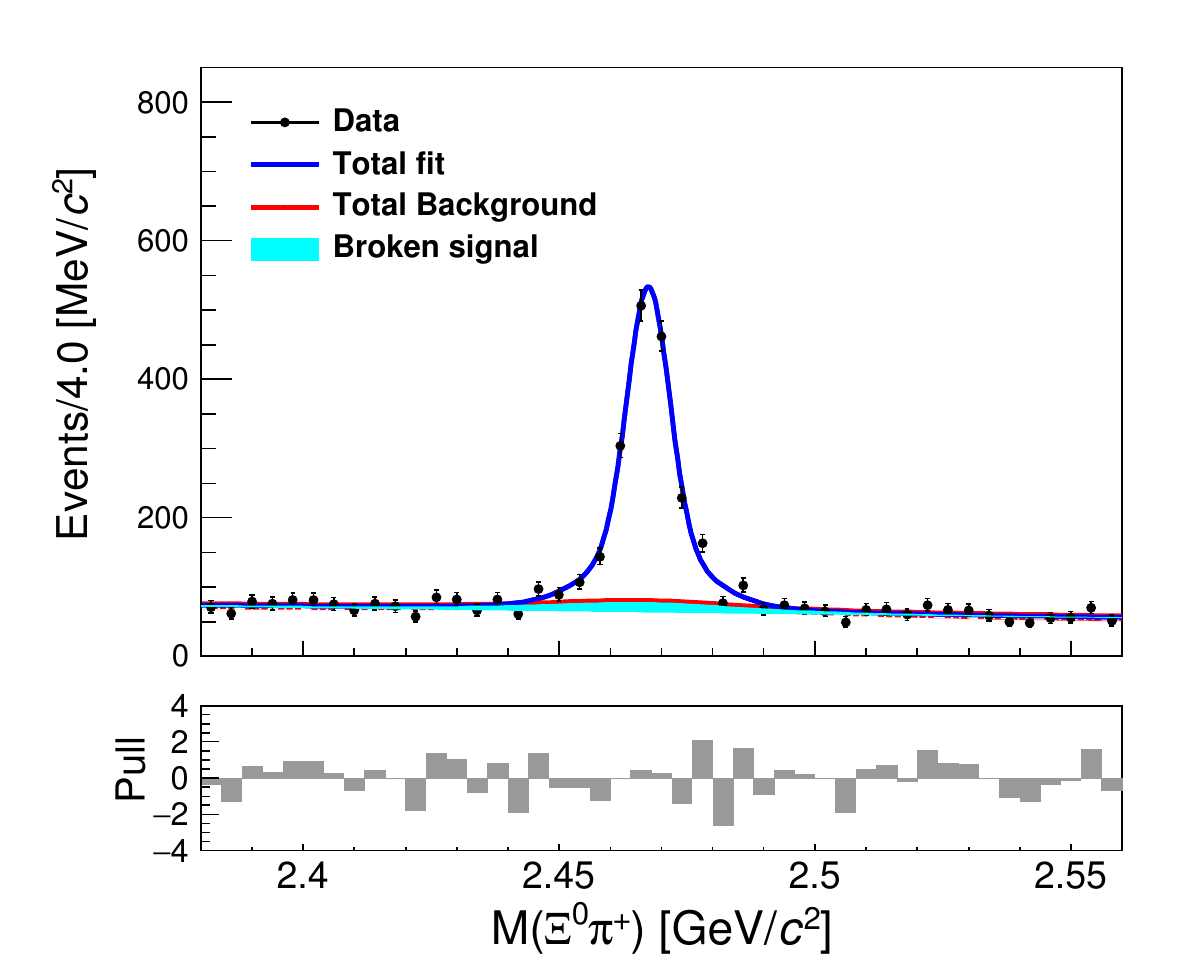}	\put(-60,130){\bf (d)}
	\put(-355,155){\scriptsize Belle} \put(-330,155){\scriptsize $\int \lum dt$ = 983.0~fb$^{-1}$}
	\put(-155,155){\scriptsize Belle~II} \put(-120,155){\scriptsize $\int \lum dt$ = 427.9~fb$^{-1}$}
\vspace{0.5cm}
	\includegraphics[width=7.0cm]{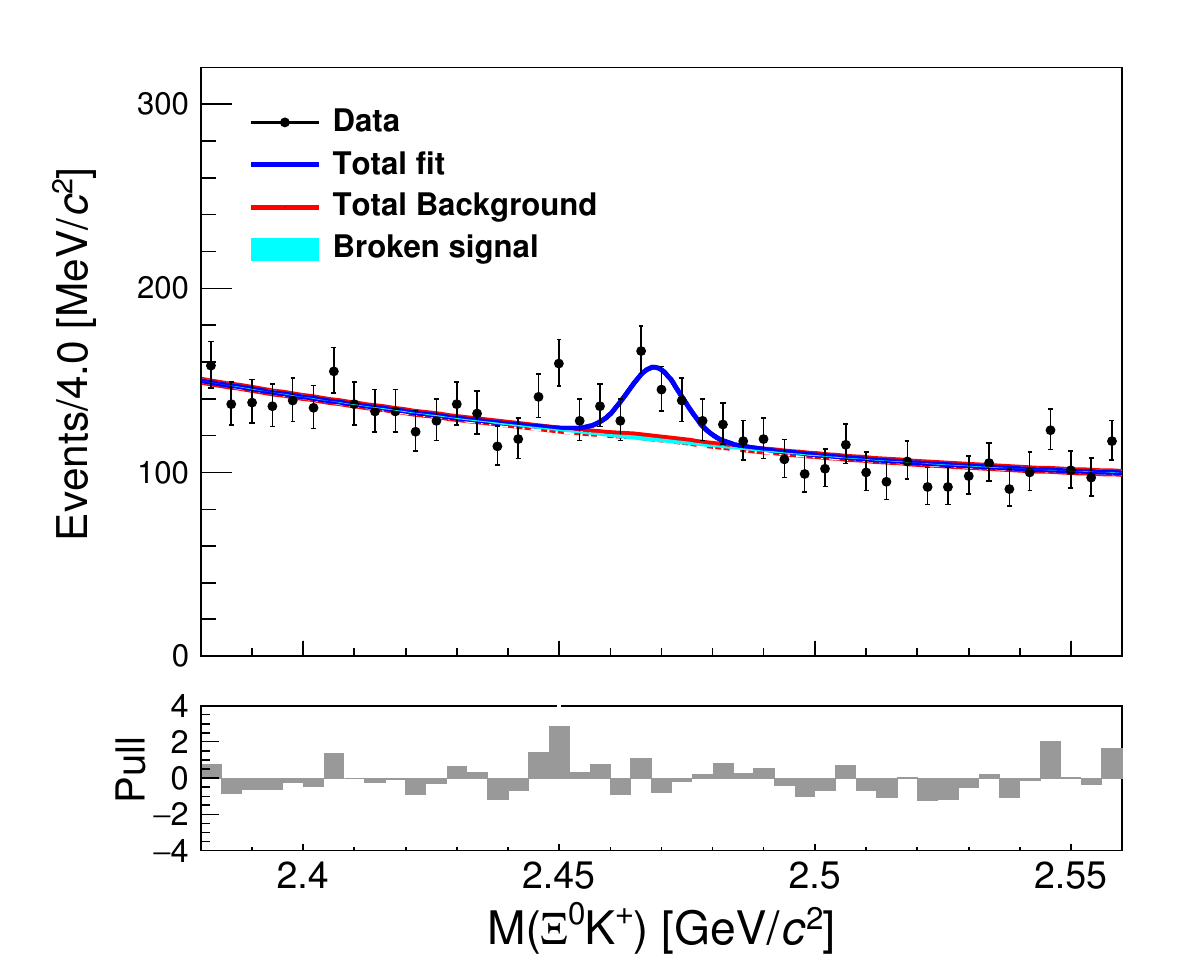}
    \put(-60,130){\bf (e)}	
	\includegraphics[width=7.0cm]{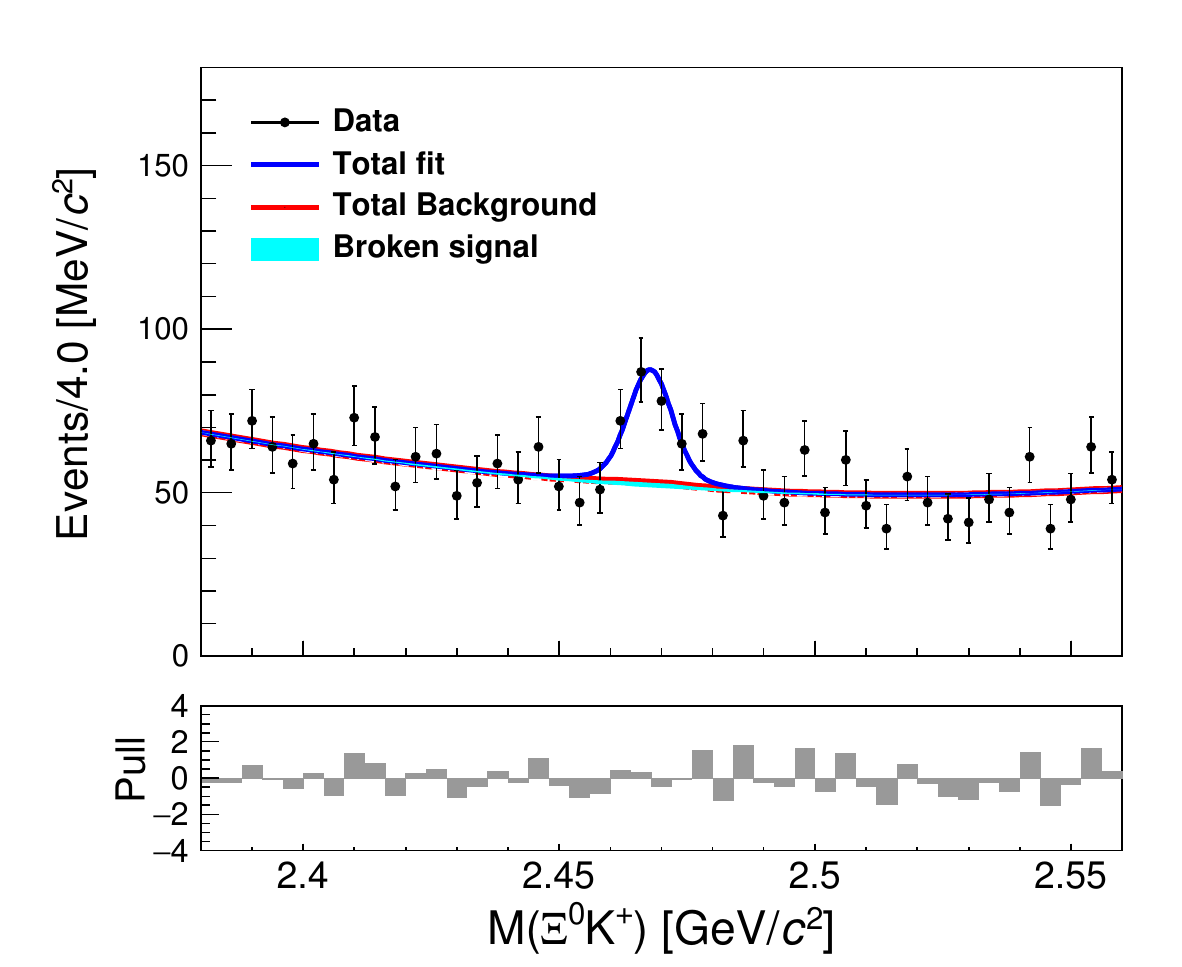}	
    \put(-60,130){\bf (f)}
	\put(-355,155){\scriptsize Belle} \put(-330,155){\scriptsize $\int \lum dt$ = 983.0~fb$^{-1}$}
	\put(-155,155){\scriptsize Belle~II} \put(-120,155){\scriptsize $\int \lum dt$ = 427.9~fb$^{-1}$}
	\caption{Invariant mass distributions of $\xicp$ candidates from (a, b) $\xicp\to\sgks$, (c, d) $\xicp\to\xipi$, and (e, f) $\xicp\to\xikp$ decays reconstructed in (left) Belle and (right) Belle~II data. The black dots with error bars show the distribution from the data. The solid blue curves and red curves show the total fit and total backgrounds, respectively. The cyan areas show the broken-signal component. The $\chi^2/{\rm ndf}$ values of the fit are 0.75 (0.88), 0.83 (1.10), and 0.82 (0.83) for $\xicp\to\sgks$, $\xicp\to\xipi$, and $\xicp\to\xikp$ at Belle (Belle II), respectively.}
\label{mxicp-data}
\end{figure}

Distributions of $M(\sgks)$, $M({\xipi})$, and $M({\xikp})$ of $\xicp$ candidates reconstructed in data are shown in figure~\ref{mxicp-data}. 
These invariant mass distributions are used to extract $\xicp$ signal yields from an unbinned EML fit.
We study the background of $M(\sgks)$ using simulation~\cite{topoana} and $\sg$ sidebands, and those of $M(\xipi)$ and $M(\xikp)$ using simulation~\cite{topoana} and $\xiz$ sidebands.
In addition to a flat combinatorial component, there is a peaking background component, referred to as broken-signal, dominated by random photons not associated with the signal decay, which is described by a non-parametric kernel estimation PDF~\cite{rookeyspdf} $({\mathcal F}_{\rm brk}$) obtained from MC simulation.
The total fit function ($\mathcal F$) includes, therefore, terms for the signal (${\mathcal F}_{\rm sig}$), ${\mathcal F}_{\rm brk}$, and smooth background (${\mathcal F}_{\rm bkg}$) contributions:
\begin{equation}
\setlength\abovedisplayskip{6pt}
\setlength\belowdisplayskip{6pt}
{\mathcal F}=n_{\rm sig}{\mathcal F}_{\rm sig}+n_{\rm brk}{\mathcal F}_{\rm brk}+n_{\rm bkg}{\mathcal F}_{\rm bkg},
\end{equation}
where $n_{\rm sig}$, $n_{\rm brk}$, and $n_{\rm bkg}$ are the number of $\xicp$ signal events, broken-signal events, and smooth background events, respectively.
The values of $n_{\rm sig}$ and $n_{\rm bkg}$ are free in the fit. The value of $n_{\rm brk}$ is not taken into account when calculating the reconstruction efficiency and signal extraction. For $\xicp\to\sgks$ and $\xicp\to\xikp$ modes, the ratios of $n_{\rm brk}$ to $n_{\rm sig}$ are fixed to the fractions from MC simulations and are 15.4\% (15.9\%) and 18.0\% (18.5\%), respectively, at Belle (Belle II). Due to the clear $\xicp$ signal for the $\xicp\to\xipi$ mode, the ratio is free. The fitting result for the ratio is consistent with the MC simulation.
The signal PDFs in $\xicp\to\sgks$ and $\xicp\to\xikp$ decays are modeled using the sum of two Gaussians with a common mean, convolved with a Gaussian function to take into account the difference in mass resolution between data and MC simulation. Here, the sum of two Gaussians is fixed to the MC simulation; we use the $\xicp\to\xipi$ as the control mode and convolve the $\xicp$ signal shapes with a Gaussian function to account for the convolved Gaussian width in the $\xicp\to\sgks$ and $\xicp\to\xikp$ signal PDFs.
For the $\xicp \to \Xi^{0} \pip$, the signal shape of $\xicp$ candidates is modeled using the sum of two Gaussians with a common mean, where the fractions and parameters are free.
The ${\mathcal F}_{\rm bkg}$ contribution is parameterized by a second-order polynomial for each mode, with all the parameters free in the fit. 
A validation of the fit through MC simulation confirms that the results are unbiased and the uncertainties follow a Gaussian distribution.
The reconstruction efficiencies and fit results for each mode are listed in table~\ref{Nxicp}.

\begin{table}
\vspace{0.2cm}
\centering
\renewcommand\arraystretch{1.2}
\footnotesize
\begin{tabular}{|l c c c c|}
		\hline
Mode & fitted yield (Belle) & $\varepsilon_{\rm Belle}$ (\%) & fitted yield (Belle~II) & $\varepsilon_{\rm Belle~II}$ (\%) \\
\hline
$\xicp\to\xipipi$ & $(487\pm4)\times10^2$ & \hspace{0.6mm}$11.37\pm 0.02$ & $(196\pm2)\times10^2$ & \hspace{0.6mm}$10.98\pm 0.02$ \\
$\xicp\to\sgks$ & ~~$288\pm 41$  & ~~$1.93\pm 0.02$ & $182\pm 31$ & ~~$2.48\pm 0.02$ \\
$\xicp\to\xipi$ & \hspace{0.4mm}$2782\pm 74$ & ~~$2.68\pm 0.02$ & \hspace{0.6mm}$1469\pm 40$~~ & ~~$3.22\pm 0.03$ \\
$\xicp\to\xikp$ & ~~$138\pm 31$ & ~~$2.25\pm 0.03$ & $100\pm 20$ & ~~$2.71\pm 0.02$ \\
		\hline
\end{tabular}
\caption{Observed $\xicp$ signal yields, statistical significances, and reconstruction efficiencies for the studied modes. Uncertainties are statistical only.}\label{Nxicp}
\end{table}

The reconstruction efficiencies for signal modes in Belle II are larger than those in Belle due to improved photon reconstruction stemming from timing improvements in the ECL readout electronics.
The statistical significances for $\xicp\to\sgks$, $\xicp\to\xipi$, and $\xicp\to\xikp$ are $7.4\sigma$ ($6.2\sigma$), >$10.0\sigma$ (>$10.0\sigma$), and $4.7\sigma$ ($5.5\sigma$) in Belle (Belle II), respectively. 
The significances are estimated using $-2 \ln({\mathcal L}_0/{\mathcal L}_{\rm max})$, where ${\mathcal L}_0$ and ${\mathcal L}_{\rm max}$ are the values of the likelihood without and with the signal component~\cite{significance}, respectively.
The difference in the number of degrees of freedom in the fit are 1 for $\xicp\to\sgks$ and $\xicp\to\xikp$ modes, and 5 for $\xicp\to\xipi$, and are taken into account to estimate the statistical significance.
Alternative fits to the $\xikp$ mass spectra are performed: (1) changing the fit range by 10\%, (2) changing the order of polynomial for the smooth background, (3) floating the ratio of $n_{\rm brk}$ to $n_{\rm sig}$. The significance for the $\xicp\to\xikp$ mode is larger than $4.5\sigma$ and $5.1\sigma$ for Belle and Belle II, respectively, in all cases.

The ratios of branching fractions to the normalization mode $\xicp\to\xipipi$ are calculated via
\begin{equation}\label{bf4}
\begin{split}
& \frac{\BR(\xicp\to\sgks)}{\BR(\xicp\to\xipipi)}= \frac{N_{\sgks}\times\varepsilon_{\xipipi}}{\varepsilon_{\sgks}\times N_{\xipipi}}\times\frac{\BR(\xim\to\Lambda\pim)\mathcal{B}( \Lambda \to p\pi^{-})}{\BR(\Sigma^{+} \to p\pi^{0})\BR(K_{S}^{0} \to \pip\pi^{-})\BR(\piz\to\gamma\gamma)},\\
& \frac{\BR(\xicp\to\xipi)}{\BR(\xicp\to\xipipi)}= \frac{N_{\xipi}\times \varepsilon_{\xipipi}}{\varepsilon_{\xipi}\times N_{\xipipi}}\times\frac{\BR(\xim\to\Lambda\pim)}{\BR(\xiz\to\Lambda\piz)\BR(\piz\to\gamma\gamma)}, \\
& \frac{\BR(\xicp\to\xiz\kp)}{\BR(\xicp\to\xipipi)}= \frac{N_{\xikp}\times\varepsilon_{\xipipi}}{\varepsilon_{\xikp}\times N_{\xipipi}}\times\frac{\BR(\xim\to\Lambda\pim)}{\BR(\xiz\to\Lambda\piz)\BR(\piz\to\gamma\gamma)}.
\end{split}
\end{equation}
Here, $N_{\sgks}$, $N_{\xipi}$, $N_{\xikp}$, and $N_{\xipipi}$ are the $\xicp$ yields resulting from the fits; $\varepsilon_{\sgks}$, $\varepsilon_{\xipi}$, $\varepsilon_{\xikp}$, and $\varepsilon_{\xipipi}$ are the corresponding reconstruction efficiencies; and the branching fractions are taken from the PDG world averages~\cite{pdg}. The calculated branching fraction ratios are summarized in table~\ref{BRxicp}.
Meanwhile, the branching fraction ratio $\BR(\xicp\to\xiz\kp)/\BR(\xicp\to\xipi)$ is calculated by 
\begin{equation}\label{bf5}
\begin{split}
& \frac{\BR(\xicp\to\xiz\kp)}{\BR(\xicp\to\xipi)}= \frac{N_{\xikp}\times\varepsilon_{\xipi}}{\varepsilon_{\xikp}\times N_{\xipi}}.
\end{split}
\end{equation}
We combine the Belle and Belle II branching fraction ratios and uncertainties using the formulas in ref.~\cite{combine},
\begin{equation}\label{eq:combineBF}
\begin{split}
r=\frac{r_1\sigma_2^2+r_2\sigma_1^2}{\sigma_1^2+\sigma_2^2+(r_1-r_2)^2\epsilon_r^2}, \\
\sigma=\sqrt{\frac{\sigma_1^2\sigma_2^2+(r_1^2\sigma_2^2+r_2^2\sigma_1^2)\epsilon_r^2}{\sigma_1^2+\sigma_2^2+(r_1-r_2)^2\epsilon_r^2}},
\end{split}
\end{equation}
where $r_i$, $\sigma_i$, and $\epsilon_r$ are the branching fraction ratio, uncorrelated uncertainty, and relative correlated systematic uncertainty of the branching fraction ratio from each data sample, respectively. All the uncorrelated and correlated uncertainties are listed in table~\ref{BRsyst}. The systematic uncertainties are discussed in Section 6.
The combined branching fraction ratios are summarized in table~\ref{BRxicp}, where the first uncertainty is statistical and the second systematic.
\begin{table}
	\vspace{0.2cm}
	\centering
    \renewcommand\arraystretch{1.2}
	\scriptsize  
	\begin{tabular}{|l c c c|}
	\hline
	Mode & Belle & Belle II & Combined \\
	\hline
$\BR(\xicp\to\sgks)/\BR(\xicp\to\xipipi)$  & $0.063\pm 0.009\pm 0.004$ & $0.075\pm 0.013\pm 0.006$ & $0.067\pm 0.007\pm 0.003$ \\
$\BR(\xicp\to\xipi)/\BR(\xicp\to\xipipi)$  & $0.246\pm 0.007\pm 0.011$ & $0.259\pm 0.007\pm 0.015$ & $ 0.251\pm 0.005\pm 0.010$ \\
$\BR(\xicp\to\xikp)/\BR(\xicp\to\xipipi)$  & $0.015\pm 0.003\pm 0.001$ & $0.021\pm 0.004\pm 0.002$ & $0.017\pm 0.003\pm 0.001$ \\
$\BR(\xicp\to\xikp)/\BR(\xicp\to\xipi)$    & $0.061\pm 0.014\pm 0.004$ & $0.081\pm 0.016\pm 0.005$ & $0.068\pm 0.010\pm 0.004$ \\
		\hline
	\end{tabular}
\caption{Branching fraction ratios of $\xicp\to\sgks$, $\xipi$, and $\xikp$ decays relative to $\xicp\to\xipipi$ and the ratio of $\BR(\xicp\to\xikp)$ and $\BR(\xicp\to\xipi)$. The first uncertainty is statistical and the second systematic.}\label{BRxicp}
\end{table}

\section{Systematic uncertainties}

There are several sources of systematic uncertainties in the measurements of the branching fraction ratios, including those associated with efficiency, the branching fraction of the intermediate state, and the fit procedure. Note that the uncertainties from efficiency-related 
sources and the branching fraction of the intermediate state partially cancel when taking the ratio to the normalization mode. Table~\ref{BRsyst} summarizes the systematic uncertainties, with the total uncertainty calculated as the quadratic sum of the uncertainties from each source.
For the branching fraction ratio $\BR(\xicp\to\xikp)/\BR(\xicp\to\xipi)$, the uncertainties related to $\xiz$ reconstruction and $\pip/ K^+$ tracking cancel, so that the remaining uncertainties come from detection efficiency-related sources (PID and MC sample size) and the fit procedure (fit uncertainty and background shape).

\begin{table}
	\centering
	\vspace{0.2cm}
	\renewcommand\arraystretch{1.5}
	\footnotesize
	\begin{tabular}{|lcccccc|}
    	\hline 
        \rule{0pt}{18pt}
		\multirow{2}{*}{Source}  & \multicolumn{2}{c}{\large $\frac{\BR(\xicp\to\sgks)}{\BR(\xicp\to\xipipi)}$} & \multicolumn{2}{c}{ \large $\frac{\BR(\xicp\to\xipi)}{\BR(\xicp\to\xipipi)}$}  & \multicolumn{2}{c|}{\large $\frac{\BR(\xicp\to\xikp)}{\BR(\xicp\to\xipipi)}$} \\
		& Belle & Belle~II& Belle & Belle~II& Belle & Belle~II\\
		\hline
		Tracking 						& 0.7  & 0.7  & 1.4  & 1.4  & 1.4 & 1.4   \\
		  PID 				              & 0.1  & 0.2  & 0.1  & 0.1  & 0.1 & 0.2   \\
		$\piz$ reconstruction 			& 2.2  & 4.2  & 2.4  & 4.3  & 2.3 & 4.3   \\
		  $\ks$ reconstruction			  & 0.8  & 2.3  & -    & -    & -   & -     \\
		  $\Lambda$ reconstruction		  & 0.5  & 0.7  & -    & -    & -   & -     \\
        Mass resolution                 & 1.4   & 1.6 & 0.4  & 0.6  & 1.1 & 1.4   \\
        MC sample size                    & 1.0  & 1.0  & 1.0  & 1.0  & 1.0 & 1.0   \\ 
        Dalitz efficiency-correction    & 1.5  & 1.7  & 1.5  & 1.7  & 1.5 & 1.7   \\
        Fit uncertainty                 & 3.7  & 4.8  & 0.9  & 1.0  & 5.3 & 4.3   \\
        Background shape     	        & 1.5  & 1.5  & 2.4  & 2.4  & 2.2 & 2.2   \\
        Intermediate states $\BR$       & 1.0  & 1.0  & 0.1  & 0.1  & 0.1 & 0.1   \\
		\hline
		Total 			                & 5.4  & 7.6  & 4.3  & 5.7  & 6.7  & 7.1  \\
		\hline
\end{tabular}
	\caption{Relative systematic uncertainties (\%) on the results of branching fraction ratios. The uncertainties in the last two rows, due to intermediate branching fractions and background shape, are common to Belle and Belle II; the other uncertainties are independent. Since the $\Lambda\to p\pim$ decay is reconstructed in $\xicp\to\xipi$ and $\xicp\to\xikp$, the $\BR(\Lambda\to p\pim)$ uncertainty and the uncertainty due to the $\Lambda\to p\pim$ reconstruction efficiency cancel in the ratios of $\frac{\BR(\xicp\to\xipi)}{\BR(\xicp\to\xipipi)}$ and $\frac{\BR(\xicp\to\xikp)}{\BR(\xicp\to\xipipi)}$.}\label{BRsyst}
\end{table}

The detection efficiencies determined from the simulations are corrected by the multiplicative data-to-simulation ratios determined from data control samples, and the uncertainties on the correction factors are taken as systematic uncertainties~\cite{xicp1bf2024}.
The correction factors and uncertainties include those from track-finding efficiency, obtained from the control samples of $D^{*+}\to D^0(\to K_S^0 \pi^+ \pi^- )\pi^+$ at Belle and $\bar B^0\to D^{*+}(\to D^{0}\pi^+)\pi^-$ and $e^+e^-\to\tau^+\tau^-$ at Belle II.
By weighting the momentum distributions of the charged tracks, we include the corresponding efficiency correction factors and the systematic uncertainties of 0.35\% and 0.33\% per track for Belle and Belle II, respectively. 
Furthermore, the tracking uncertainties of $\pip$ and $K^+$ mesons, originating from the $\xicp$, do not cancel in the ratios $\frac{\BR(\xicp\to\xipi)}{\BR(\xicp\to\xipipi)}$ and $\frac{\BR(\xicp\to\xikp)}{\BR(\xicp\to\xipipi)}$ due to the tighter momentum requirements. The systematic uncertainties for other charged tracks cancel since the normalization and signal modes have similar distributions.
At Belle, the PID uncertainties for charged pion, kaon, and proton are studied using $D^{*+} \to  D^0(\to K^-\pip)\pip$ and $\Lambda \to p \pim$~\cite{BellePID1} control samples, respectively.
At Belle~II, the corresponding PID uncertainties are obtained using $D^{*+} \to  D^0(\to K^- \pip)\pip$, $\ks \to \pip\pim$, and $\Lambda \to p \pim$~\cite{PIDBelle2} control samples, respectively.
The uncertainties of $\piz$ reconstruction are obtained from the $\tau\to\pi^-\pi^0\nu_\tau$ control sample at Belle and the $D^0\to K^-\pi^+\piz$ control sample at Belle II. 
The uncertainty associated with the mass windows of the $\pi^0$ is calculated from the data-simulation difference on the fraction of the fitted $\piz$ signal yield in the $\piz$ signal region over that in the total region.
This uncertainty is added in quadrature with the contribution from $\piz$ reconstruction.
The $K_{S}^{0}$ reconstruction uncertainties are obtained from the $D^{*+} \to  D^0(\to\ks \pip\pim)\pip$ control samples at Belle and Belle II.
Since we applied a decay length selection for $K_{S}^{0}$ candidates in the Belle II data, the uncertainty is calculated from the data-simulation difference on the fraction of the fitted $K_{S}^{0}$ signal yield in the $K_{S}^{0}$ signal region divided by that in the total region.
This uncertainty is added in quadrature with the uncertainty related to the $K_{S}^{0}$ reconstruction.
The uncertainties of $\Lambda$ reconstruction are obtained from the $\Lambda\to p\pi^-$ and $\Lambda^+_{c}\to\Lambda(\to p\pim)\pip$ control samples at Belle and Belle II, respectively.
The uncertainties of the intermediate particle ($\Sigma^+$ and $\xiz$) signal region selections are calculated from the data-simulation difference on the fraction of the $\Sigma^+$ and $\xiz$ signal yield in the $\Sigma^+$ and $\xiz$ signal region divided by that in the total signal region, respectively.
The systematic uncertainty due to the limited MC simulation sample sizes is calculated using a binomial uncertainty estimate. 
For the reference mode $\xicp\to \xipipi$, the detection efficiency is corrected across the Dalitz plot. 
The selected $\xicp$ sideband regions may influence the efficiency. We enlarge the $\xicp$ sideband regions by a factor of two, and the deviation in efficiency compared to the nominal value is taken as systematic uncertainty.

The uncertainties due to the fit procedure are determined by taking the differences between the $\xicp\to\sgks$ and $\xicp\to\xikp$ signal yields in the nominal fits and the signal yields in fits with the following modifications: 
$(1)$ changing the fit range by $10\%$, $(2)$ changing the order of polynomial for the smooth background, $(3)$ floating the ratio of $n_{\rm brk}$ to $n_{\rm sig}$, and $(4)$ changing the convolved Gaussian width by $\pm 1\sigma$.
Only the fit range and background shape are considered as the sources of uncertainty for the $\xicp\to\xiz\pip$ mode.
The order of the polynomial for the background shape is the same for the two data samples, and the corresponding uncertainty is extracted from a simultaneous fit for the $\xicp$ signal yield in the Belle and Belle II data. We treat the background shape uncertainty as a separate systematic uncertainty.
The uncertainties associated with changing the broken-signal PDF smoothed by $\texttt{RooKeysPdf}$ to $\texttt{RooHistPdf}$~\cite{rookeyspdf,roohistpdf} are smaller than 0.1\% and are neglected.
The total systematic uncertainty is obtained by adding the contributions from each source in quadrature. 
Since charmed baryons are produced inclusively at Belle and Belle II, the possible polarization effect is small~\cite{polar}. We also checked the $\xicp$ angular distributions between data and MC simulations and found that they are consistent.
In addition, we weight the signal MC samples based on the efficiency-corrected $x_p$ distributions of the normalization mode from data to improve agreement between data and MC simulations. The efficiency-corrected $x_p$ distribution is determined by fitting the $M(\xim\pip\pip)$ distribution in each $x_p$ bin of data while incorporating the efficiency in each bin~\cite{xicp1bf2024}.
Thus, the systematic uncertainty associated with the model of signal MC generation can be neglected. For the normalization mode $\xicp\to\xipipi$, the uncertainties are determined by changing the fit range and the order of the background polynomial. These uncertainties are added in quadrature for the corresponding branching fraction ratio measurements.

The systematic uncertainties due to the intermediate branching fractions are taken to be the uncertainties on the world-average values~\cite{pdg} and treated as correlated uncertainties, which are common to Belle and Belle~II.
For the measurement of $\frac{\mathcal{B}(\Xi_{c}^{+}\to \Sigma^{+}K_{S}^{0})}{\mathcal{B}(\Xi_{c}^{+}\to \Xi^{-}\pi^{+}\pip)}$, the uncertainties from 
$\mathcal{B}( \Sigma^{+} \to p\pi^{0})$, $\mathcal{B}( K_{S}^{0} \to \pip\pi^{-})$, $\mathcal{B}( \pi^{0} \to \gamma\gamma)$, $\mathcal{B}( \Xi^{-} \to \Lambda\pi^{-})$, and $\mathcal{B}( \Lambda \to p\pi^{-})$ are 0.58\%, 0.07\%, 0.03\%, 0.04\%, and 0.78\%~\cite{pdg}, which are added in quadrature as the total uncertainty from branching fractions of intermediate states.
For the measurements of $\frac{\mathcal{B}(\xicp \to \Xi^{0}\pi^{+})}{\mathcal{B}(\Xi_{c}^{+}\to \Xi^{-}\pi^{+}\pip)}$ and $\frac{\mathcal{B}(\xicp \to \Xi^{0}K^{+})}{\mathcal{B}(\Xi_{c}^{+}\to \Xi^{-}\pi^{+}\pip)}$, the uncertainties from $\mathcal{B}( \Xi^{-} \to \Lambda\pi^{-})$, $\mathcal{B}( \Xi^{0} \to \Lambda\pi^{0})$, and $\mathcal{B}( \pi^{0} \to \gamma\gamma)$ are 0.04\%, 0.01\%, and 0.03\%~\cite{pdg}, respectively.
The 44.8\% uncertainty on $\BR(\xicp\to\xipipi)$~\cite{xicpabsbf2019} is treated as an independent systematic uncertainty in the measurement of the absolute branching fractions.

Adding the contributions from each source in quadrature in table~\ref{BRsyst}, the total systematic uncertainties are $5.4\%$~($7.6\%$), $4.3\%$~($5.7\%$), $6.7\%$~($7.1\%$), and $6.5\%$~($6.2\%$) for $\frac{\BR(\xicp\to\sgks)}{\BR(\xicp\to\xipipi)}$, $\frac{\BR(\xicp\to\xipi)}{\BR(\xicp\to\xipipi)}$, $\frac{\BR(\xicp\to\xikp)}{\BR(\xicp\to\xipipi)}$, and $\frac{\BR(\xicp\to\xikp)}{\BR(\xicp\to\xipi)}$ at Belle (Belle II), respectively.

\section{Summary and discussion}

We present measurements of the branching fractions of $\Xi_c^{+}$ decays into $\Sigma^{+}K_{S}^{0}$, $\Xi^{0}\pi^{+}$, and $\Xi^{0}K^{+}$ final states, using 983.0~fb$^{-1}$ and 427.9~fb$^{-1}$ data samples collected by the Belle and Belle~II experiments, respectively. By using the decay $\Xi_c^{+} \to \Xi^{-} \pip \pip$ as a reference mode, relative branching fractions are measured to be
\begin{equation} \label{rbf1}\frac{\mathcal{B}(\Xi_{c}^{+} \to \Sigma^{+}K_{S}^{0})}{\mathcal{B}(\Xi_{c}^{+}\to \Xi^{-}\pi^{+}\pip)} = 0.067\pm 0.007  \pm 0.003 ,\end{equation}
\begin{equation} \label{rbf2}\frac{\mathcal{B}(\Xi_c^{+} \to \Xi^{0}\pi^{+})}{\mathcal{B}(\Xi_{c}^{+}\to \Xi^{-}\pi^{+}\pip)}
= 0.251\pm 0.005\pm 0.010,\end{equation}
and
\begin{equation} \label{rbf3}\frac{\mathcal{B}(\Xi_c^{+} \to \Xi^{0}K^{+})}{\mathcal{B}(\Xi_{c}^{+}\to \Xi^{-}\pi^{+}\pip)} = 0.017\pm 0.003 \pm 0.001,\end{equation}
where the first uncertainties are statistical and the second are systematic.
The ratio of ${\mathcal B}(\Xi_{c}^{+}\to\Xi^{0}K^{+})$/${\mathcal B}(\Xi_{c}^{+}\to\Xi^{0}\pi^{+})$ is measured to be $0.068\pm 0.010\pm 0.004$.
Taking $\mathcal{B}(\Xi_c^+ \to \Xi^{-} \pip \pip)$ = ($2.9\pm 1.3$)\%~\cite{pdg}, the absolute branching fractions are measured to be
\begin{equation}\mathcal{B}(\Xi_{c}^+ \to \Sigma^{+}K_{S}^{0}) = (0.194\pm 0.021  \pm 0.009 \pm 0.087 )\%,\end{equation}
\begin{equation}~\mathcal{B}(\Xi_c^+ \to \Xi^{0}\pi^{+}) = (0.728 \pm 0.014 \pm 0.027  \pm 0.326 )\%,\end{equation}
and
\begin{equation}\mathcal{B}(\Xi_c^+ \to \Xi^{0}K^{+}) = (0.049 \pm 0.007  \pm 0.003  \pm 0.022)\%,\end{equation}
where the third uncertainties are from $\BR(\Xi_c^{+} \to \Xi^{-} \pip \pip)$.

\begin{figure}[!htbp]
\centering
\begin{minipage}[c]{0.6\textwidth}
\includegraphics[width=\textwidth]{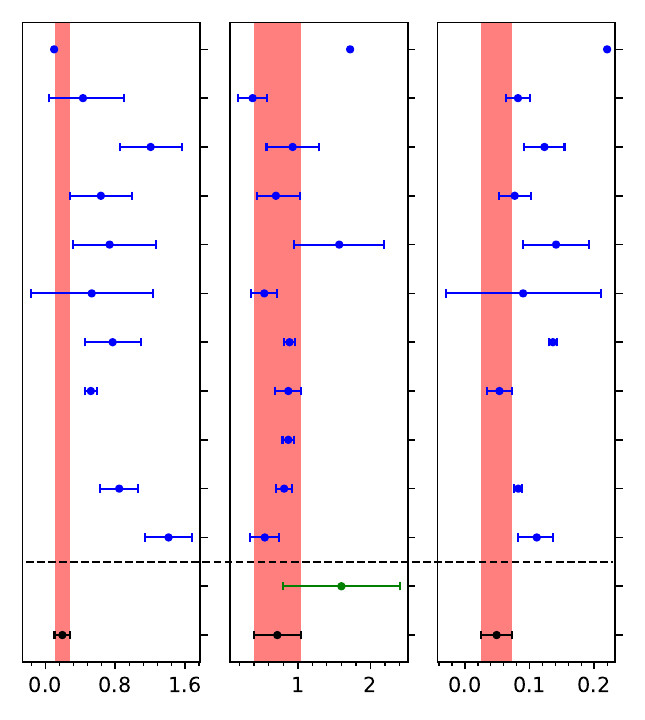}\\[-\baselineskip]
\text{\footnotesize \hspace{0.1cm} $\BR(\Xi_c^+ \to \Sigma^+ K_{S}^0)(\%)$ \hspace{0.1cm} $\BR(\Xi_c^+ \to \Xi^0 \pi^+)(\%)$ \hspace{0.1cm} $\BR(\Xi_c^+ \to \Xi^0 K^+)(\%)$}\\
\end{minipage}
\begin{minipage}[c]{0.3\textwidth}
\footnotesize
 Zou et al.~\cite{theory10poleca2020}\\[2.6mm]
 Geng et al.~\cite{theory8su3f2019}\\[2.6mm]
 Zhao et al.~\cite{theory9su3f2020}\\[2.6mm]
 Hsiao et al. (I)~\cite{theory12su3f2022}\\[2.6mm]
 Hsiao et al. (II)~\cite{theory12su3f2022}\\[2.6mm]
 Huang et al.~\cite{theory11su3f2022}\\[2.6mm]
 Xing et al.~\cite{theory14su3f2023}\\[2.6mm]
 Liu et al. (I)~\cite{theory100su3f2024}\\[2.6mm]
 Liu et al. (II)~\cite{theory100su3f2024}\\[2.6mm]
 Zhong et al. (I)~\cite{theory13su3f2023}\\[2.6mm]
 Zhong et al. (II)~\cite{theory13su3f2023}\\[2.6mm]
 CLEO Collaboration~\cite{exp1}\\[2.6mm]
 Belle and Belle~II\\
 combined measurements\\\vspace{3mm}
\end{minipage}
\caption{The comparisons of the measured $\BR(\Xi_c^+ \to \Sigma^+ K_{S}^0)$, $\BR(\Xi_c^+ \to \Xi^0 \pi^+)$, and $\BR(\Xi_c^+ \to \Xi^0 K^+)$ with theoretical predictions~\cite{theory10poleca2020,theory8su3f2019,theory9su3f2020,theory11su3f2022,theory12su3f2022,theory14su3f2023,theory100su3f2024,theory13su3f2023}. The blue dots with error bars represent theoretical predictions. The green dot with an error bar indicates the result from the CLEO Collaboration~\cite{exp1}, and the black dots with error bars correspond to the results of this study. The horizontal dashed line distinguishes the theoretical predictions from the experimental results. The red vertical bands cover a $1\sigma$ region corresponding to the measurements presented in this work. Dots with error bars represent central values and their uncertainties; those without indicate a lack of theoretical uncertainty. Missing dots signify the absence of theoretical predictions or experimental results for that decay mode. For refs.~\cite{theory12su3f2022,theory13su3f2023}, (I) indicates the predicted value based on the ${\rm SU(3)}_{f}$ symmetry, while (II) takes into account the breaking of ${\rm SU(3)}_{f}$ symmetry. For the ref.~\cite{theory100su3f2024}, (I) and (II) represent the predicted values derived from low-lying pole and general pole scenarios along with ${\rm SU(3)}_{f}$, respectively.} \label{comparsion}
\end{figure}

Figure~\ref{comparsion} shows the comparison of measurements of the branching fractions with theoretical calculations as mentioned in Section 1. Two predictions utilizing SU(3)$_{f}$ symmetry with the irreducible approach method~\cite{theory8su3f2019,theory11su3f2022} agree well with our measurements across all decay channels. 
Most theoretical predictions, considering their uncertainties, are consistent with the measured central result within 2.4$\sigma$ for the SCS decay $\Xi_c^+ \to \xikp$. However, the predictions based on the dynamical model~\cite{theory10poleca2020} and the SU(3)$_{f}$ method~\cite{theory14su3f2023,theory13su3f2023} differ significantly from the measured central value.
The measured absolute branching fraction of $\Xi_c^+ \to \Sigma^+ K_S^0$ is lower than the central values predicted by most theoretical papers.
Our measurement of $\BR(\xicp\to\xipi)$ is consistent with the result from the CLEO Collaboration and has better precision~\cite{exp1}.
The ratios (\ref{rbf1}), (\ref{rbf2}), and (\ref{rbf3}) are independent of the $\Xi_c^+$ absolute branching fraction scale and may serve as valuable benchmarks for comparison with theoretical models.

\acknowledgments
This work, based on data collected using the Belle II detector, which was built and commissioned prior to March 2019, and data collected using the Belle detector, which was operated until June 2010,
was supported by
Higher Education and Science Committee of the Republic of Armenia Grant No.~23LCG-1C011;
Australian Research Council and Research Grants
No.~DP200101792, 
No.~DP210101900, 
No.~DP210102831, 
No.~DE220100462, 
No.~LE210100098, 
and
No.~LE230100085; 
Austrian Federal Ministry of Education, Science and Research,
Austrian Science Fund (FWF) Grants
DOI:~10.55776/P34529,
DOI:~10.55776/J4731,
DOI:~10.55776/J4625,
DOI:~10.55776/M3153,
and
DOI:~10.55776/PAT1836324,
and
Horizon 2020 ERC Starting Grant No.~947006 ``InterLeptons'';
Natural Sciences and Engineering Research Council of Canada, Compute Canada and CANARIE;
National Key R\&D Program of China under Contract No.~2022YFA1601903 and No.~2024YFA1610503,
National Natural Science Foundation of China and Research Grants
No.~11575017,
No.~11761141009,
No.~11705209,
No.~11975076,
No.~12135005,
No.~12150004,
No.~12161141008,
No.~12475093,
and
No.~12175041,
and Shandong Provincial Natural Science Foundation Project~ZR2022JQ02;
the Czech Science Foundation Grant No.~22-18469S 
and
Charles University Grant Agency project No.~246122;
European Research Council, Seventh Framework PIEF-GA-2013-622527,
Horizon 2020 ERC-Advanced Grants No.~267104 and No.~884719,
Horizon 2020 ERC-Consolidator Grant No.~819127,
Horizon 2020 Marie Sklodowska-Curie Grant Agreement No.~700525 ``NIOBE''
and
No.~101026516,
and
Horizon 2020 Marie Sklodowska-Curie RISE project JENNIFER2 Grant Agreement No.~822070 (European grants);
L'Institut National de Physique Nucl\'{e}aire et de Physique des Particules (IN2P3) du CNRS
and
L'Agence Nationale de la Recherche (ANR) under Grant No.~ANR-21-CE31-0009 (France);
BMBF, DFG, HGF, MPG, and AvH Foundation (Germany);
Department of Atomic Energy under Project Identification No.~RTI 4002,
Department of Science and Technology,
and
UPES SEED funding programs
No.~UPES/R\&D-SEED-INFRA/17052023/01 and
No.~UPES/R\&D-SOE/20062022/06 (India);
Israel Science Foundation Grant No.~2476/17,
U.S.-Israel Binational Science Foundation Grant No.~2016113, and
Israel Ministry of Science Grant No.~3-16543;
Istituto Nazionale di Fisica Nucleare and the Research Grants BELLE2,
and
the ICSC – Centro Nazionale di Ricerca in High Performance Computing, Big Data and Quantum Computing, funded by European Union – NextGenerationEU;
Japan Society for the Promotion of Science, Grant-in-Aid for Scientific Research Grants
No.~16H03968,
No.~16H03993,
No.~16H06492,
No.~16K05323,
No.~17H01133,
No.~17H05405,
No.~18K03621,
No.~18H03710,
No.~18H05226,
No.~19H00682, 
No.~20H05850,
No.~20H05858,
No.~22H00144,
No.~22K14056,
No.~22K21347,
No.~23H05433,
No.~26220706,
and
No.~26400255,
and
the Ministry of Education, Culture, Sports, Science, and Technology (MEXT) of Japan;  
National Research Foundation (NRF) of Korea Grants
No.~2016R1-D1A1B-02012900,
No.~2018R1-A6A1A-06024970,
No.~2021R1-A6A1A-03043957,
No.~2021R1-F1A-1060423,
No.~2021R1-F1A-1064008,
No.~2022R1-A2C-1003993,
No.~2022R1-A2C-1092335,
No.~RS-2023-00208693,
No.~RS-2024-00354342
and
No.~RS-2022-00197659,
Radiation Science Research Institute,
Foreign Large-Size Research Facility Application Supporting project,
the Global Science Experimental Data Hub Center, the Korea Institute of
Science and Technology Information (K24L2M1C4)
and
KREONET/GLORIAD;
Universiti Malaya RU grant, Akademi Sains Malaysia, and Ministry of Education Malaysia;
Frontiers of Science Program Contracts
No.~FOINS-296,
No.~CB-221329,
No.~CB-236394,
No.~CB-254409,
and
No.~CB-180023, and SEP-CINVESTAV Research Grant No.~237 (Mexico);
the Polish Ministry of Science and Higher Education and the National Science Center;
the Ministry of Science and Higher Education of the Russian Federation
and
the HSE University Basic Research Program, Moscow;
University of Tabuk Research Grants
No.~S-0256-1438 and No.~S-0280-1439 (Saudi Arabia), and
Researchers Supporting Project number (RSPD2025R873), King Saud University, Riyadh,
Saudi Arabia;
Slovenian Research Agency and Research Grants
No.~J1-9124
and
No.~P1-0135;
Ikerbasque, Basque Foundation for Science,
the State Agency for Research of the Spanish Ministry of Science and Innovation through Grant No. PID2022-136510NB-C33,
Agencia Estatal de Investigacion, Spain
Grant No.~RYC2020-029875-I
and
Generalitat Valenciana, Spain
Grant No.~CIDEGENT/2018/020;
The Swiss National Science Foundation;
The Knut and Alice Wallenberg Foundation (Sweden), Contracts No.~2021.0174 and No.~2021.0299;
National Science and Technology Council,
and
Ministry of Education (Taiwan);
Thailand Center of Excellence in Physics;
TUBITAK ULAKBIM (Turkey);
National Research Foundation of Ukraine, Project No.~2020.02/0257,
and
Ministry of Education and Science of Ukraine;
the U.S. National Science Foundation and Research Grants
No.~PHY-1913789 
and
No.~PHY-2111604, 
and the U.S. Department of Energy and Research Awards
No.~DE-AC06-76RLO1830, 
No.~DE-SC0007983, 
No.~DE-SC0009824, 
No.~DE-SC0009973, 
No.~DE-SC0010007, 
No.~DE-SC0010073, 
No.~DE-SC0010118, 
No.~DE-SC0010504, 
No.~DE-SC0011784, 
No.~DE-SC0012704, 
No.~DE-SC0019230, 
No.~DE-SC0021274, 
No.~DE-SC0021616, 
No.~DE-SC0022350, 
No.~DE-SC0023470; 
and
the Vietnam Academy of Science and Technology (VAST) under Grants
No.~NVCC.05.12/22-23
and
No.~DL0000.02/24-25.

These acknowledgements are not to be interpreted as an endorsement of any statement made
by any of our institutes, funding agencies, governments, or their representatives.

We thank the SuperKEKB team for delivering high-luminosity collisions;
the KEK cryogenics group for the efficient operation of the detector solenoid magnet and IBBelle on site;
the KEK Computer Research Center for on-site computing support; the NII for SINET6 network support;
and the raw-data centers hosted by BNL, DESY, GridKa, IN2P3, INFN, 
and the University of Victoria.

\renewcommand{\baselinestretch}{1.2}

\begin{appendices}
\appendix
\end{appendices}

\end{document}